\documentclass[showpacs,prd,amsmath,amssymb,twocolumn]{revtex4}
\usepackage{graphicx}
\usepackage{epsfig}
\usepackage{graphics}
\usepackage{amsmath}
\usepackage{dcolumn}
\usepackage{bm}
\usepackage{algorithmicx}

\begin{document}

\title{Comment on ``Quasinormal modes of Schwarzschild anti de Sitter black holes:
Electromagnetic and gravitational perturbations"}


\author{A. B. Pavan}
\email{alan@unifei.edu.br}
\affiliation{Instituto de F\'{i}sica e Qu\'{i}mica, \\
Universidade Federal de Itajub\'{a}}
\author{Rodrigo Silva Lima}
\email{rodlima@unifei.edu.br}
\affiliation{Instituto de Matem\'{a}tica  e Computac\~ao, \\
Universidade Federal de Itajub\'{a}}

\author{Lu\'{i}s Filipe de Almeida Roque}
\email{filipe_roque1@hotmail.com}

\affiliation{ Instituto de Engenharia Mec\^anica , \\
Universidade Federal de Itajub\'{a}}

\begin{abstract}
The quasinormal modes of the electromagnetic and gravitational perturbation on Schwarzschild-AdS black hole calculated in \cite{Cardoso} has been revisited. Although the equations of motion are correct some frequencies calculated previously by the authors are not. We present the new values of quasinormal modes and discuss the possible sources of problems and implications on the conclusions presented.
\end{abstract}

\pacs{04.70.-s, 04.25.Nx, 04.30.Nk}
\maketitle

In \cite{Cardoso} the authors have calculated the quasinormal modes (QNM) of electromagnetic and gravitational perturbation on Schwarzschild-AdS black holes. They start the paper deriving the equations which drive the evolution of both perturbations and discuss some analytical properties related to the stability of black hole when submitted to both perturbations. After that, the numerical procedure using power-series methods \cite{Horowitz} was described and implemented to find the QNM and then the results obtained were presented and discussed.
The most part of discussions of results performed by the authors were organized taking into account the size of black holes: small $(r_{+}\ll1)$, intermediate $(r_{+}\sim 1)$ and large $(r_{+}\gg1)$ and the parity of perturbation: even and odd. Here we will follow the same structure adopted for theirs.
As the authors we have programmed a notebook of the \emph{Mathematica} program to calculate the QNM using an adapted version of the algorithm developed by Cardoso and shared in his homepage \footnote{http://centra.tecnico.ulisboa.pt/$\sim$vitor/?page=ringdown}. Later, revisiting the problem and results presented by authors, one can observed that our numerical results were qualitative and quantitatively different to those presented previously in \cite{Cardoso}. So we decided to write this letter to share our results.

\section{Numerical Results and Discussions} \label{numex}

We start presenting the QNM's that we have found and later discussing the possible reasons for the differences among that values found by Cardoso and Lemos.

\subsection{Electromagnetic perturbation}

Firstly we calculated the QNM's for electromagnetic perturbation with $\ell=1,2$. For $\ell=1$ we found the same values that Cardoso and Lemos, however for $\ell=2$ the values shown are different. This difference caught our attention especially because it was dependent on the angular quantum number $\ell$ and it seems to not be dependent of the algorithm programmed. However, in this case, the algorithm was exactly the problem.

\begin{table}[htbp]
\renewcommand{\arraystretch}{1.3}
\tabcolsep 5pt
\caption{Lowest QNM of electromagnetic perturbations for $l=1,2$.}
\label{tabelectro}
\begin{tabular}{ccccccc}
\hline
\hline
 & & $\hspace{-1.5cm}\ell=1$  &  & &$\hspace{-1.5cm}\ell=2$  & \\
\hline
$r_+$ & $\omega_R $ & $\omega_I$ &  $N$ &$\omega_R $ & $\omega_I$ &  $N$\\
\hline
0.8& $-2.175 $  & $-1.287   $  & 40 & $-3.224 $    & $-0.996  $  & 60 \\
1  & $-2.163 $  & $-1.699   $  & 40 & $-3.223 $    & $-1.384  $  & 40 \\
5  & $\sim 0 $  & $-8.795   $  & 40 & $-3.090 $    & $-9.822  $  & 40 \\
10 & $\sim 0 $  & $ -15.506 $  & 40 & $\sim 0 $    & $-16.623 $  & 40 \\
50 & $\sim 0 $  & $-75.096  $  & 40 & $\sim 0 $    & $-75.269 $  & 40 \\
100& $\sim 0 $  & $-150.048 $  & 40 & $\sim 0 $    & $-150.134$  & 40 \\
\hline
\hline
\end{tabular}
\hspace{0.1cm}
\end{table}

We claim that the reason for this incorrect values lies on the term $\ell(\ell+1)$ of the effective potential in the algorithm developed by Cardoso and Lemos. When they were to program the effective potential in the notebook they missed the factor of $\ell$. This argument is illustrated by the equations below
\begin{eqnarray}
V_{\ell}&=& f(r)\ \frac{\ell\ (\ell+1)}{r^2},\\
\nonumber\\
\tilde{V}_{\ell}&=&f(r)\ \frac{(\ell+1)}{r^2},
\end{eqnarray}
where the effective potential $V$ is the potential correctly programmed and $\tilde{V}$ is not well-programmed potential. As one can see, it explains why the QNM for $\ell=1$ are in agreement and for $\ell=2$ are not once that for $\ell=1$, $V_{1}=\tilde{V}_{1}$, but for $\ell\geq2$, $V_{\ell\geq2}\neq\tilde{V}_{\ell\geq2}$. To test our hypothesis we calculated the QNM with $\tilde{V}$ and could recover all values for $\ell=2$ presented in \cite{Cardoso}.

Thus, our results for the lowest QNM of electromagnetic perturbation for $\ell=1,2$ are compiled and presented in Table \ref{tabelectro} where the frequency is written as $\omega=\omega_{R}+ i \omega_{I}$. We have included in Table \ref{tabelectro} the maximum number of iterations $N$ necessary to the convergence of the solutions for each value of QNM. Cardoso has pointed out that in \cite{Kokkotas}, Berti and Kokkotas have presented the same table of values for these QNM. However we argue that they did not discuss the origin of the differences between the values of the original work and those found by them.

At first sight the QNM of electromagnetic perturbation for large black holes are not strongly dependent on small values of $\ell$. However, for $\ell > 10$ the imaginary part changes substantially with $\ell$. For intermediate and small black holes this dependency is explicit in Table \ref{tabelectro}. Probably, the authors did not calculate the QNM for sufficiently high values of $\ell$ to find this dependency. The interpretation of the authors about the scaling of imaginary part of QNM with $r_+$ (for large black holes), in the AdS/CFT context, is not affected by the new results.

\subsection{Gravitational perturbation}

For the gravitational perturbation, we have found discrepancies in the values of QNM to intermediate and small black holes. The most important differences appear in the QNM for gravitational odd perturbation and are concentrated in Tables III and IV shown in \cite{Cardoso}. In this case, one cannot identify any discrepancies between the algorithms.

Comparing the algorithm used by the authors with ours, we could see that the frequencies were calculated with the command \emph{FindRoot} while we used the command \emph{NSolve}. We believe that it can be the reason why we have obtained different values for QNM especially for small black holes. It is because the command \emph{FindRoot} depends strongly on the initial complex value $\omega_{0}$ to find only one root, whereas the command \emph{NSolve} looking for all roots of the polynomial equation in a defined range of values, allowing us to select the appropriated value in the set of values found.
The most important differences in QNM that we found appeared for $r_{+}=0.5$ and $r_{+}=1$. The ``algebrically special value" mentioned could not be found with our algorithm.

\begin{table}[htbp]
\renewcommand{\arraystretch}{1.3}
\tabcolsep 5pt
\caption{Lowest QNM of gravitational odd perturbations for $l=2$. Here $\omega_0$ is the fundamental mode and $\omega_1$ is the first overtone}
\label{tabgravoddl2}
\begin{tabular}{ccccccc}
\hline
\hline
& & $\hspace{-1.5cm}\omega_0$&& &$\hspace{-1.5cm}\omega_1$& \\
\hline
$r_+$ & $\omega_R $ & $\omega_I$ &  $N$ &$\omega_R $ & $\omega_I$ &  $N$\\
\hline
0.5&$-3.037$     & $-0.719 $  & 70 & $-4.34283$    & $-2.0998   $ & 120 \\
1  &$-3.033$     & $-2.404 $  & 40 & $-4.96041$    & $-4.8981   $ & 50 \\
2  &$\sim 0$     & $-0.7285$  & 40 & $-4.44748$    & $-5.2583   $ & 40 \\
5  &$\sim 0$     & $-0.2703$  & 40 & $-9.57712$    & $-13.294   $ & 40 \\
10 &$\sim 0$     & $-0.1338$  & 40 & $-18.6618$    & $-26.626   $ & 40 \\
50 &$\sim 0$     & $-0.0267$  & 40 & $-92.5047$    & $-133.190  $ & 40 \\
100&$\sim 0$     & $-0.0133$  & 40 & $-184.958$    & $-266.384  $ & 40 \\
\hline
\hline
\end{tabular}
\hspace{0.1cm}
\end{table}

Furthermore, the definition of the fundamental mode is not the standard in this case. In general the lowest QNM, i.e., the fundamental mode, is associated to the lowest value of the imaginary part $\omega_I$ because it is the last mode to decay since $\Psi(t,r)=\psi(r)\ e^{- i\omega_{R} t}\ e^{\omega_{I}t}$. However, in Tables III and IV shown in \cite{Cardoso} the authors applied a different definition relating the lowest mode to $\omega_{R}$. For example in Table III, the lowest QNM for $r_+=0.5$ had $\omega=(0 -i\ 6.4)$ while the second lowest mode had $\omega=(3.037-i\ 0.72) $ which is smaller than previous value. Here we have obtained values that are consistent with standard definition.

In Tables \ref{tabgravoddl2} and \ref{tabgravoddl3} we list the lowest QNM of the gravitational odd perturbations for $\ell=2$ and $\ell=3$.

\begin{table}[htbp]
\renewcommand{\arraystretch}{1.3}
\tabcolsep 5pt
\caption{Lowest QNM of gravitational odd perturbations for $l=3$. Here $\omega_0$ is the fundamental mode and $\omega_1$ is the first overtone}
\label{tabgravoddl3}
\begin{tabular}{ccccccc}
\hline
\hline
& & $\hspace{-1.5cm}\omega_0$&  && $\hspace{-1.5cm}\omega_1$&  \\
\hline
$r_+$ & $\omega_R $ & $\omega_I$ &  $N$ &$\omega_R $ & $\omega_I$ &  $N$\\
\hline
0.5& $-4.185$    & $-0.389$ & 100&  $-5.308$     & $-1.516$&120 \\
1  & $-3.849$    & $-1.639$ & 60 &  $-5.238$   & $-4.185$& 60\\
2  & $\sim 0 $   & $-2.189$ & 40 &  $-4.615$   & $-5.080$& 40\\
5  & $\sim 0 $   & $-0.690$ & 40 &  $-9.725$   & $-13.247$& 40\\
10 & $\sim 0 $   & $-0.336$ & 40 &  $-18.743$   & $-26.603$& 40\\
50 & $\sim 0 $   & $-0.067$ & 40 &  $-92.521$   & $-133.185$& 40\\
100& $\sim 0 $   & $-0.033$ & 40 &  $-184.967$   & $-266.382$& 40\\
\hline
\hline
\end{tabular}
\hspace{0.1cm}
\end{table}

For the gravitational even perturbation we have not found any discrepancies in the values of QNM. The Table \ref{tabgraveven} shows the values that we found. It is presented only to check the consistency of our algorithm.

\begin{table}[htbp]
\renewcommand{\arraystretch}{1.3}
\tabcolsep 4.5pt
\caption{Lowest QNM of gravitational even perturbations for $l=2,3$.}
\label{tabgraveven}
\begin{tabular}{ccccccc}
\hline
\hline
& & $\hspace{-1.5cm}\ell=2$&   &  & $\hspace{-1.5cm}\ell=3$ &   \\
\hline
$r_+$ & $\omega_R $ & $\omega_I$ &  $N$ &$\omega_R $ & $\omega_I$ &  $N$\\
\hline
1  & $-3.018   $   & $-1.584$     & 70 & $-3.910   $    & $-1.390    $  & 70 \\
2  & $-4.546   $   & $-3.974$     & 70 & $-4.597   $    & $-3.299    $  & 50 \\
5  & $-9.832   $   & $-12.650$    & 60 & $-10.219  $    & $-11.642   $  & 60 \\
10 & $-18.806  $   & $-26.301$    & 45 & $-19.091  $    & $-25.789   $  & 50 \\
50 & $-92.535  $   & $-133.125$   & 45 & $-92.596  $    & $-133.022  $  & 50 \\
100& $-184.974 $   & $-266.351$   & 45 & $-185.005 $    & $-266.300  $  & 45 \\
\hline
\hline
\end{tabular}
\end{table}

Another subject mentioned in \cite{Cardoso} was the influence of the higher values of angular quantum number $\ell$ in the QNM. In our analysis we have found a different behaviour. In general, for $\ell\geq10$, the values of frequencies depend on $\ell$. A detailed analysis about this subject is under investigation actually and will appear in \cite{Alan}

\section{Conclusions} \label{final}

In summary some values of QNM for electromagnetic and gravitational perturbations on Schwarzschild-AdS black hole have been revised. The most part of the new results concerns to intermediate and small black holes. Small problems in the algorithm programmed by the authors were identified, fixed and discussed. We have discussed the definition of fundamental mode (lowest mode) used to construct Tables III and IV in \cite{Cardoso} and presented the standard definition. Finally, four tables with new values of QNM were presented and discussed.

\acknowledgments

The authors would like to thank FAPEMIG, FAPESP and CNPq by the financial support.

\end{document}